# Mapping the Information Journey: Unveiling the Documentation Experience of Software Developers in China[1]

Zhijun Gao
Peking University
zhijun.gao@pku.edu.cn

Jiangying Wang
Peking University
EdmodoW@126.com

Meina Wang
Ant Group
wcqwmn@126.com

**Abstract**

**Purpose:** Understanding the developers as well as their behaviors in using technical documentation is essential to the creation of high quality developer documentation. In this research, we explore Chinese developers' characteristics in terms of the use of technical documentation, their information journey, and their expectations for quality documentation.

**Method:** We interviewed 25 software developers and surveyed 177 participants. The design of the survey was based on the preliminary findings of the interviews. Adopting traditional user research approaches, persona and user journey mapping, we drew typical personas and information journeys of the participants based on the qualitative interview data and the quantitative survey results.

**Results:** We identified several common characteristics and differences between junior developers and senior developers in terms of the use of technical documentation, categorized as personality, learning habits, and working habits. We also found that the information journey of both two groups of developers roughly follows these four stages: Exploration, Understanding, Practice, and Application. Hence, we drew two personas and two information journey maps to represent the two groups of developers. The content, organization, and maintenance are the top three documentation dimensions that developers care about.

**Conclusion:** We suggested organizing documentation content based on developers' information journey, adapting documentation to the needs of developers at different levels, and prioritizing the content, organization, and maintenance of documentation.

**Keywords:** Developer documentation, user research, persona, information journey

## Introduction

With the development of cloud computing, artificial intelligence, and other Internet technologies, many developer-oriented Internet products have emerged in China. Developer documentation, which is defined as "a special type of documentation designed to help software developers create or improve a system", is an integral part of these products, because the quality of the documentation directly affects the developer experience and the credibility of the products. This group of documentation includes source code comments, tutorials, reference documentation for APIs, design documentation, etc (Lethbridge, Singer, & Forward, 2003). In addition, some researchers believe that in addition to official documents, User Generated Content (UGC) also belongs to developer documents, including open source projects, technical blogs, and technical Q&A platforms (Li, Xing, Peng, & Zhao, 2013).

[1] This research received funding from the Peking University-Ant Group collaboration under the Ministry of Education's Industry-University-Research Initiative, focusing on the project titled "Developer Documentation Experience Design and Evaluation."

Technical writers of developer documentation attach great importance to the quality of the documentation. Delivering efficient documentation and minimizing barriers for developers are their daily goals. To help technical writers better write and evaluate developer documentation, it is necessary to understand the characteristics of the developers, how they interact with the documentation, and what they expect from the documentation. As a classic method to improve user experience and usability, user research has been widely used in consumer-facing products, but it is rarely used in studies on developers. A few user studies have been aiming to understand developers' searching behaviors (Li, Xing, Peng, & Zhao, 2013; Masudur Rahman et al., 2018), reading behaviors (Meng et al., 2019), and debugging behaviors (Beller, Spruit, Spinellis, & Zaidman, 2018), yet there has been little research on how they search and use developer documentation. Therefore, we focus our research on developers' interaction with documentation. We are interested in the behaviors of developers in the process of seeking and using documentation as an aid for their work. Since developer documentation is a specific type of technical information, the interaction between developers and developer documentation is an information process. We intend to understand the whole information process, thus we reference the concept of information journey mapping, a visualization method that addresses the initiation of information needs and people's behaviors in understanding and using the information to meet their needs (Howard, 2014). The information journey map could shed light on developers' information needs, preferences, behaviors, etc.

Current studies on software developers are within a broad and general geographical scope. As this study is situated in China, a fast-growing software market with its characteristics, we consider the characteristics of Chinese developers. Existing studies on Chinese developers are mainly from commercial organizations such as the Chinese Software Developer Network (CSDN), which provide limited information about the demographic features and technology preferences of these developers. Some academic research focuses on the roles and main activities of Chinese developers (Zhang & Xie, 2018), as well as their working time (Zhang et al., 2020), but they lack a deep exploration of their habits and behaviors. In this study, we focus on the characteristics of Chinese developers in terms of the use of technical documentation. Considering that the mental models and habits of junior and intermediate developers may differ, we also wanted to compare the differences between these two groups of developers.

The study is aimed at exploring the following three fundamental questions related to Chinese software developers.

**RQ1:** What are the characteristics of Chinese developers in terms of the use of technical documentation? Is there any difference between junior and senior developers?

**RQ2:** What is the information journey of Chinese developers?

**RQ3:** What do Chinese developers expect from the documentation?

This research enriches the understanding of developers by presenting the findings from semi-structured interviews with 25 software developers and a survey among 177 developers across major cities in China where the Internet industry is developed. To facilitate readers' use of our research results, we drew the personas and information journey maps of the developers based on the qualitative and quantitative data. This study also adds to the body of literature by identifying developers' expectations for high-quality technical documentation.

**Literature review**

**Difference between developers and end users**

In a broad sense, developers are a subgroup of users, because developers are users of various development tools including programming languages, integrated development environments (IDEs), Software Development Kits (SDKs), and Application Programming Interfaces (APIs) (Mikkonen & Programming, 2016). In the field of software, we define end users as the ones that are supposed to use an application and developers as the ones that develop an application to fulfill the needs of end users. Both end users and developers use technical documentation to assist in completing tasks or using products, but with different motivations and cognitive styles.

Developers have stronger incentives to read documentation than end users. Documentation may be optional for end users. For example, a user may learn to use an e-reader without referring to its instructions. For developers, however, documentation is an essential support. Software development is a sophisticated activity that demands the professional skills of developers. The advent of numerous development tools like APIs have made it easier for software developers to eliminate the need to code from scratch (Meng, Steinhardt, & Schubert, 2019). However, due to the professionalism and complexity of the tools themselves, developers cannot use these tools without learning. Unlike end-users, developers have a strong dependency on technical documentation. They often consult technical documentation as an aid in understanding and using complex codes, components, and services throughout the process of development (Lewko & Parton, 2021). Milewski (2007) found that browsing documentation has been one of the day-to-day activities of software developers. Previous research showed that documentation was one of the starting points in the process of comprehending software (Roehm, Tiarks, Koschke & Maalej, 2012).

End users are a broad group whose members' cognitive levels are difficult to measure. Developers, on the other hand, are a relatively concentrated group with higher cognitive abilities. Most computer software development jobs require at least a bachelor's degree in computer science or software engineering. A survey showed that over 66% of the software developers had bachelor's degrees or above (Vailshery, 2022). One research that compared the cognitive styles of software developers and common testees showed that developers exhibited a more holistic and imagery cognitive style. This means that developers tend to process information from a holistic, global perspective and prefer to represent information through images. Another research revealed that software developers exhibited a technology-oriented cognitive style. In other words, when facing business requirements, they prefer to think about how to implement requirements technically rather than thinking from end users' viewpoints.

To sum up, developers are a more professional group and have a stronger demand for technical documentation. Therefore, it is necessary to distinguish developers from end users and conduct a more in-depth study on developers and developer documentation.

**Characteristics of developers in terms of documentation use**

There is generally a lack of research examining the characteristics of developers in terms of the use of technical documentation. Some studies explored the types of documentation that developers used. Through content analysis, two latest studies analyzed the questions and mailing lists related to documents submitted by developers on technology platforms such as Stack Overflow and GitHub. They identified the documentation issues that developers are concerned about and the document types commonly used by developers (Aghajani et al., 2020; Aghajani et al., 2019). Video tutorials seem to be a new type of technical documentation for developers. A study on development tutorials reveals that developers tend to rely on video tutorials to learn new content but text tutorials for checking information when both types of tutorials are available (Käfer, Kulesz, & Wagner, 2016). Watching videos can be more time-consuming than reading text documentation, but it is a more intuitive way to understand abstract knowledge. Clarke (2007) explored how developers used API documentation and found that there were three typical strategies: systematic, opportunistic, and pragmatic strategies. Developers using the first strategy usually try to have a thorough understanding of the API before dealing with details of the development task. Some developers are used to starting to code first and then searching for information in the API reference as needed. This is called the opportunistic strategy. The third type, the pragmatic approach, is a combination of systematic and opportunistic strategies.

Good knowledge of the characteristics of developers, especially their needs, preferences, habits, and behaviors, can be meaningful for the design of developer documentation. The current literature on this issue is still underexplored, so we hope this study can fill some gaps.

**Information behaviors of developers**

Information behavior can be divided into two sub-categories, which are information searching behavior and information use behavior. Previous research on developers' behavior mainly focuses on behaviors in the process of searching and obtaining information. Some research focuses on what developers search for when they search the internet. Li, Xing, Peng, & Zhao (2013) explored the type of information searched by developers, the scenarios of search, and the process model of search through experiments. They divided the search process of developers into four stages: search, screening, integration, and verification. Rose (2006) analyzed the characteristics of user search behavior, including the diversity of targets, the context of search, and the iterative nature of search tasks, and proposed methods to redesign the user search interface. Sadowski, Stolee, & Elbaum (2015) used the questionnaire survey and logged analysis methods to study developers' information behavior. They found that developers frequently searched for answers to questions about how to use API documentation, the function of the code, the reasons for errors, etc. A study by Xia et al. (2017) surveyed 235 software engineers from 21 countries and identified several search tasks that developers found challenging. There are also studies investigating how developers search for and use information. Ko, Myers, Coblenz, & Aung (2006) explored the process of developers understanding unfamiliar code and development environments through user experiments, revealing the problems encountered in the search process. Meng et al. (2019) found through observation that developers often adopt an opportunistic strategy when encountering problems, that is, relying more on intuition to try various searched possible solutions, rather than carefully reviewing the instructions in the documentation.

Some of the findings of existing research can be used as a reference for the design of the interview outline of this research. However, existing research lacks a detailed description of the process of developers interacting with technical documents or technical information. Although two studies summarize the process of how developers search for information, there is a lack of analysis of developers' needs, preferences, and experiences in this process (Ko et al., 2006; Li et al., 2013). And these factors are important factors to consider when designing user-centered developer documentation. In addition, the process model of the aforementioned research only focuses on the search process of developers, lacking a deep exploration of how they use various information and resources to complete their development tasks. To describe the interaction process between developers and documents as completely as possible, this research will draw the typical information journey maps of developers.

**Developers' expectations for quality documentation**

In terms of what is quality developer documentation, previous research has not given a systematic answer. Documentation quality assessment and design overlook the involvement of developers. Evaluating documentation quality is of great significance for providing a better experience for developers. However, so far most of the existing studies have focused on evaluation criteria related to the documentation themselves, such as readability, understandability, and usability (Rama, Kak, & Experience, 2015). Research on documentation quality from the perspective of users, namely the developers, has been limited. The research focuses of previous studies on developer documentation were limited to automatic assessment tools (Dautovic, Plosch, & Saft, 2011), metrics-based evaluating models (Aversano, Guardabascio, & Tortorella, 2017), and expert opinions (Garousi, Garousi, Moussavi, Ruhe, & Smith, 2013). Developers' participation was rarely discussed in such studies. One research summarized the documentation issues that are relevant to developers through a survey. These issues were categorized into classic quality dimensions such as correctness, completeness, and up-to-dateness, which are similar to what was defined in *Developing Quality Technical Information* (Carey et a.l, 2014). Some technical writers (Bhatti et al., 2021) have published guidelines trying to standardize the writing and evaluation of developers. However, due to a lack of empirical research, it is unknown how well these guidelines will work efficiently.

**Summary of literature review**

To conclude, our review of existing literature showed that the current understanding of developers in the perspective of developer documentation is still limited. First, user research methods are rarely used in this field. Some studies discuss issues related to developer documentation but do not take developers as research subjects. Second, there are so many types of developer documentation that single research may concentrate on only one type of documentation. For example, many studies research how developers use API documentation. Therefore, we lack a general understanding of how developers use various types of documents and resources to complete development tasks. Third, there is no comprehensive research on developers' behaviors, challenges, and needs in the whole process of interacting with multiple documentation. We lack a general picture of the whole information journey of developers.

## Methods

We used the semi-structured interview approach because it provides an in-depth understanding of the participants' daily practices and helps discover some common patterns in their behaviors. To test the findings from the interviews, we surveyed a larger population. The design of the questionnaire was based on preliminary findings from the interviews.

### Interviews
### Interview guide and procedure

We designed an interview guide before conducting the interviews. The guide contained four parts. We started each interview by asking general questions about the professional background and development experience of the interviewee, and then we asked the interviewee questions related to the three research objectives. The second part was about the characteristics of developers. What needs to be clarified is that the questions used in the interviews evolved iteratively throughout the study. The questions that may confuse the interviewees were explained and clarified. For example, when asked what the characteristics of a developer are, some interviewees didn't know where to start. Thus we switched to asking about the difference between developers and people in other professions and gave some directions. The third part of the guide was intended to get insight into how developers interact with documentation in their work. We asked the participants to recall a recent experience learning a technical product (e.g. software, toolkit, frame, or programming language). The interviewees were encouraged to describe in as much detail as possible the whole process from hearing of the product to learning to use it to complete their development tasks or projects. The last part was about developers' opinions on documentation experience. We asked the developers to summarize the characteristics of their assumed good documentation. The guide is shown in Table 1.

**Table 1. The Interview Guide**

| Targets | Questions |
| --- | --- |
| Know about the basic information of the developer | Q1: Can you tell us about your professional background and development experience? |
| | Q2: How long have you been doing development work? |
| Discover the | Q3: What do you think are the differences between developers and people in other professions and |
| characteristics of developers | occupations? (e.g. personality, characteristics/learning methods/habits of learning, characteristics of work, lifestyle, etc.) |
| Discover how developers learn to use development tools | Q4: Did you learn any new technology, tools, or software recently?<br>Q5: What is the whole process from hearing of this tool to finally starting to use it like? (You can roughly divide the process into stages, such as the early, middle, and later stages.) |
| Discover what developers expect from the documentation | Q6: Can you recall examples of good or bad documentation you've used? You can tell me what's good and what's bad about them.<br>Q7: What kind of document do you think is a good technical document? Please briefly summarize several dimensions. |

Before the interviews, the participants were informed of the research goals and the interview topics. We gained their informed consent for recording the interviews and using their non-sensitive information in the research. Thirteen participants were interviewed face to face, and 12 interviews were finished through phone calls due to the COVID-19 pandemic and geographic restrictions. All the interviews were conducted in Chinese. The interview time ranged from 30 minutes to 50 minutes. All the questions in the interview guide were asked.

**Participants**

For the interviews, a total of 25 Chinese developers were recruited via WeChat groups and bulletin board systems. Each interviewee received 50 RMB as a reward. The participants were selected based on two fundamental criteria. First, they should have experience in work related to software development. Second, they should have experience in using developer documentation. To clarify, 11 of the participants were postgraduate students in computer science-related majors. They met the above two requirements because they had the relevant experience when completing course projects or internships at IT companies.

The participants consisted of 19 males and 6 females. They were between 22 and 36 years old. Two-thirds of the participants held a master's degree, and the rest held a bachelor's degree. Their development experience ranged from 1 year to 10 years, and their experience in using technical documentation was almost synchronized with their development experience. Since we aim to compare the difference between junior and senior developers, we define the former as developers with no more than five years of work experience, and the latter referred to those developers with more than five years of experience in software development. In addition to the student participants, the rest of the participants were from a wide range of specialties, including algorithm engineers, development engineers, database administrators, project managers, and technical coaches. The programming languages they often used were diverse, including JAVA, Python, C++, Rust, and C sharp.

**Data analysis**

The interview recordings were transcribed verbatim into Word documents. The Chinese version and the machine-translated version of the scripts are stored in our [GitHub repository](). Theme analysis was were employed to analyze the transcribed interview data (Braun & Clarke, 2006; ). This approach starts with identifying concepts, the words or phrases that represent an idea in the data based on the researcher's interpretations. Similar concepts are grouped to generate a higher level of abstraction, categories or themes. Qualitative analysis software ATLAS.ti was used for open coding and analysis. We first coded relevant text segments while reading through some of the interview scripts. For ample, we coded "I like to learn by doing projects" as "learn by practice". Then we developed a coding scheme consisting of codes related to the research questions to facilitate the coding process. The remaining scripts were coded following the scheme, and new codes were added to the scheme iteratively. Similar codes were then grouped into subcategories or categories to generate themes. For example, the codes "try writing down the sample codes" and "try running a small demo" were grouped into the subcategory "practice", and "practice" belonged to a higher level of category "behavior". The second author performed the open coding and the other authors reviewed the codes, subcategories, and categories. Conflicts were resolved through discussion to ensure the coding was unbiased.

**Survey**

**Instrument**

The design of the questionnaire was based on the preliminary findings from the interviews. The questionnaire had a total of 25 questions. The first ten questions were about the basic information of the respondents. The second section (Q11-Q15) was about the characteristics of developers in terms of the use of technical documentation. We asked specific questions regarding themes emerging from the interviews. For example, Q12 consisted of multiple choices related to the ways of learning: by taking offline courses or training, by watching online courses and video tutorials, by reading content from technical communities, forums, and social media, and by reading the official developer documentation. These ways of learning were summarized from the theme analysis. The third section (Q16-Q24) was related to the information journey of developers. We investigated the subcategories summarized from the interview data, which were the purpose, type of information, ways of accessing information, behaviors, and ways of interacting with information during the journey. We identified four stages in the information journey from the interview result, so we asked whether these four stages roughly match the respondent's process of learning new development tools through Q16. If the answer was "yes", the respondent was supposed to answer all questions except for Q21. If the answer was "no", the respondent would skip other questions and explain the reasons in Q21. The last section (Q25) was a scale matrix measuring the influence of different documentation features on the document usage experience of developers. The features were 23 themes we identified in the scripts corresponding to the last part of the interview guide, and developers' expectations for documentation. Moreover, two irrelevant questions for testing attention were added to the questionnaire to filter out inattentive respondents.

**Procedure**

The questionnaire was administered to participants via the online questionnaire platform wjx.com within one week. Inattentive respondents were filtered out and a total of 177 valid answers were attained. The respondent of each valid questionnaire received three RMB as a reward.

### Respondents

In the survey, a total of 177 valid answer sheets were collected. The sample consisted of 120 males and 57 females. Participants younger than 30 accounted for approximately 60% of the sample, and the rest were developers between 31 to 40 years old. Student developers accounted for 43% of the participants, and the rest were working developers. They have been working in the field of software development for 5 years on average. Their workplaces were mainly distributed in large cities of China where the information industry was thriving, such as Beijing, Shanghai, Shenzhen, Hangzhou, and Wuhan. About 62% of the participants held a bachelor's degree, and others held a master's or doctor's degree.

### Data analysis

We used quantitative methods to analyze data collected from the questionnaire survey. SPSS 26.0 was used for data analysis. We performed a descriptive analysis of the data to have an overview of the results. The Cronbach's Alpha coefficient of the documentation quality criteria matrix in the questionnaire was 0.879, which indicated that the reliability of the scale is high. The normality test showed that the questionnaire data did not conform to the normal distribution ($p<0.05$).

## Qualitative results

### Characteristics of developers

The interviewees shared many characteristics of developers. We selected those characteristics that are closely related to their daily work, especially the use of technical documentation, and organized the results in Table 2.

**Table 2. Characteristics of Developers**

| Categories | Codes | Examples |
|---|---|---|
| Personability | Open source spirit | "introverted in reality, but active in developer communities." |
| | | "very open-source and love to express themselves online through texts, audios, videos, etc." |
| | Attention to details | "pursuit of perfection" |
| | | "rigorous and attention to detail" |
| Learning habits | Strong information search ability | "strong search ability" |
| | Ability to learn quickly | "ability to accept new knowledge constantly" |
| | | "ability to learn quickly" |
| | Self-learning | "accustomed to self-study" |
| | Fragmented learning | "The knowledge learned is fragmented." |
| | | "lack of systematic learning" |
| | | "limited time to study" |
| | Learning by practicing | "learn by working on projects" |
| Working habits | High-intensity work | "long work hours" "high mental work" |

From the above analysis, we identified some common representative features related to developers' core needs for documentation. They have strong needs for information because development work requires them to constantly learn new things. The interviewees all admitted that software development is an occupation that requires fast and constant learning. The answer of one junior participant explained the reason in the interview:

> *Computer science is such a vast field that no one can learn everything well. The knowledge and technology development and iterate in a fast way. You have to have the ability to learn quickly and constantly to catch up with the technology.* (P9, Female, 1 year of experience)

Another common feature is that developers are used to self-learning and learning by practicing. The majority of the interviewees agreed that learning by doing was the more practical way because they seldom have a long period to learn things systematically. In addition, practicing the skills in projects was deemed as an effective way of becoming familiar with a product. As one developer stated:

> *When I read the documentation, I couldn't wait to test the sample codes in an IDE project whenever I had some basic understanding of a tool or a frame. I have to try it to see if it works.* (P1, Male, three years of experience)

In addition, contrary to previous stereotypes about developers, the interviewees indicated that developers were very open-sourced and willing to share and express themselves online.

We identified one difference between junior and senior developers. Most junior developers said that they preferred reading English documentation directly due to the poor quality of Chinese translation, while senior developers with 8 to 10 years of work experience preferred Chinese documentation. The preference was related to their English levels. Those who preferred reading Chinese documentation have a relatively lower level of English proficiency.

**Information journey of developers**

Since the process the learning to use a development tool can be long and full of details, during the interviews, we asked the interviewees to roughly divide the process into stages, such as the early, middle, and later stages. It turned out that in addition to the early, middle, and later stages, there was a pre-learning stage. We identified five aspects in the information journey, which were: purpose, type of information, way of accessing information, action, and key points of information design. Therefore, when analyzing the scripts of this part, we specified a coding format that consisted of three elements: stage, aspect, and code. The format was "stage-aspect-code". For example, when an interviewee talked about how he or she installed, deployed, and configured the development environment in the early stage of the journey, we coded the corresponding scripts as "early stage-action-install, deploy, and configure environment". Then all the codes were put into a two-dimensional table to form a draft of the journey. The horizontal row is the stages of the journey, and the columns are the five aspects involved in the journey.

While analyzing the interview scripts, we found that the developers demonstrated some common features in terms of the overall learning process, but there were also some differences between junior developers and senior developers. One difference between junior and senior developers lies in the way of getting started. Nine developers with less than three years of development experience mentioned that they usually got started with new technology by watching video tutorials, while almost all senior developers said that they preferred self-learning by reading text materials. They also preferred solving problems at work by themselves. For those beginning learners, documentation in the form of text is more abstract and difficult to understand, and videos provide a more intuitive effect of learning. However, watching videos can be time-consuming, which may not be suitable for senior developers who usually work on tight schedules. Reading texts may not be difficult for senior developers because they have prior knowledge. One senior developer explained:

*One can migrate what he had learned about one technology to another. For example, the grammar of several major programming languages is almost the same. If you have learned C, you are not starting from scratch when you learn Python.* (P-18, Male, 8 years of experience)

Noticing the differences, we drew two different drafts of the information journey maps for junior and senior developers respectively based on the codes. The two drafts are shown in Figure 1 and Figure 2.

| Junior Developers' Information Journey Mapping | | | | |
|---|---|---|---|---|
| Stages | Pre-learning stage | Early stage | Middle stage | Later stage |
| Purpose | 1. To finish course/research/competition/internship projects<br>2. To evaluate whether to use the product | 1. To get a general understanding of basic grammar/functions/overall architecture/algorithms/of the product | 1. To be familiar with the product<br>2. To implement small functions | 1. To implement advanced functions<br>2. To resolve specific bugs/problems<br>3. To learn about technical details |
| Type of information | 1. Internal documents<br>2. Community knowledge<br>3. Introduction/getting started document | 1. Textbooks<br>2. Video tutorials<br>3. Introduction/getting started documents<br>4. User manuals<br>5. Community knowledge (Chinese translation)<br>6. Release notes<br>7. Academic papers | 1. Code comments<br>2. How-tos/tutorials<br>3. Community knowledge<br>4. Academic paper<br>5. Codes of internal projects | 1. API reference<br>2. Function descriptions<br>3. How-tos/tutorials<br>2. FAQ/Troubleshooting<br>3. Community knowledge |
| Way of accessing information | 1. Teachers/friends/colleagues<br>2. Search engines/communities/social media<br>3. Official sites | 1. Official sites<br>2. Search engines/communities | 1. Official sites<br>2. Search engines/communities<br>3. Links of open source projects<br>4. Colleagues | 1. Official sites<br>2. Search engines/communities<br>3. Teachers/friends/colleagues |
| Action | 1. Ask teachers/friends/colleagues<br>2. Read internal documents<br>3. Search for community knowledge<br>4. Read official documents<br>5. Comparing products | 1. Watch video tutorials<br>2. Read textbooks<br>3. Read introduction/getting started documents<br>4. Read community content<br>5. Install, deploy, configure environment<br>6. Copy codes and run a small demo | 1. Refer to sample codes<br>2. Refer to existing codes written by colleagues<br>3. Try to write codes to run a small demo<br>4. Learn simple common functions<br>5. Try to implement small functions | 1. Read advanced function descriptions/task steps<br>2. Implement advanced functions<br>3. Improve existing functions<br>4. Check for specific interfaces/parameters/functions<br>5. Search answers on search engines/communities<br>6. View source code and official documents to analyze bugs<br>7. Copy and paste error messages<br>8. Consult teachers/friends/colleagues<br>9. Raise issues at GitHub/QQ groups |
| Key points of information design | SEO/community promotion | Videos/diagrams/interactive code examples/online running environment | Navigations/inner search engines/interactive code examples/online running environment/ | Navigations/inner search engines/code examples/tables/one-click copy/communities |

**Figure 1. The Draft of Junior Developers' Information Map**

| | Senior Developers' Information Journey Mapping | | | |
|---|---|---|---|---|
| Stages | Pre-learning stage | Early stage | Middle stage | Later stage |
| Purpose | 1. To finish work projects<br>2. To get a general understanding of functions/advantages of the products<br>2. To choose which product to use<br>3. To know about new trends in the industry | 1. To get a general understanding of basic concepts/logic/grammar/functions/usage/design ideas of the product | 1. To be familiar with the product<br>2. To test the feasibility of the product<br>3. To implement specific functions related to | 1. To know the underlying logic of the product<br>2. To implement specific functions related to business<br>3. To resolve specific bugs/problems |
| Type of information | 1. Introduction/getting started documents<br>2. promotional articles<br>3. Community knowledge<br>4. API reference<br>5. FAQ<br>6. Source code<br>7. Academic paper | 1. Introduction/getting started documents<br>2. How-tos/tutorials<br>3. API reference<br>4. FAQ/troubleshooting<br>5. Release notes<br>6. Academic papers<br>7. Internal documents<br>8. Community knowledge | 1. How-tos/tutorials<br>2. API reference<br>3. Source code | 1. API reference<br>2. How-tos/tutorials<br>3. FAQ/troubleshooting<br>3. Community knowledge<br>4. Source code |
| Way of accessing information | 1. Search engines/communities/social media/mailing list<br>2. Colleagues<br>3. Official sites<br>4. Conferences | 1. Official sites<br>2. Search engines/communities<br>3. Internal knowledge base | 1. Official sites<br>2. Search engines/communities | 1. Official sites<br>2. Search engines/communities/GitHub projects<br>3. Colleagues |
| Action | 1. Ask friends/colleagues<br>2. Search for product comparison information on search engines and communities<br>3. Search for best practices of other companies<br>4. Search for and read official documents/source code<br>5. Compare different products and evaluate their feasibility | 1. Read introduction/getting started documents/tutorials/API reference/FAQs/Release notes/internal documents<br>2. Search for and read community content<br>3. Install, deploy, configure environment<br>4. Copy codes and run a small demo | 1. View specific sections related to business needs<br>2. Refer to sample code<br>3. Write code to practice and implement small<br>4. Search for answers<br>5. Conduct experiments on the product<br>6. Try to implement functions related to business | 1. Read tutorials and sample codes<br>2. View specific interfaces/parameters/functions<br>3. Implement specific functions<br>4. Improve details<br>5. Check codes and debug with tools<br>6. Search answers on search engines/communities<br>7. Copy and paste error messages<br>8. View source code and official documents<br>9. Consult colleagues<br>10. Raise issues at GitHub/QQ groups/Q&A sites |
| Key points of information design | Community promotion/SEO/competitive product analysis | Navigations/inner search engines/interactive code examples/online running environment | Navigations/inner search engines | Navigations/inner search engines/code examples/tables/one-click copy/communities |

**Figure 2. The Draft of Senior Developers' Information Map**

**The pre-learning stage**

The pre-learning marks the initiation or motivation of the information behavior. For junior developers, the initiation of the learning process shows a passive pattern. Most junior developers said that they decided to learn a new tool or technology because their teachers or leaders asked them to do so. They usually heard of a new technical product from their friends or superiors. In contrast, senior developers show a more active pattern in this stage. They research widely to choose the most appropriate product for their work project. Their information sources include technical communities, social media, official sites, and academic papers. This purposeful research before using a tool or a product is called technology selection. One participant explained the reason for doing this:

> *In a team-based working environment, I have to persuade other people if I decide to use one tool or frame, so I need to know its advantages and disadvantages very well to evaluate whether it is suitable for our project.* (P17, Male, 6 years of experience)

While a junior developer may refer to only introduction documents or community knowledge to have an overall understanding of a product, senior developers read a variety of documents such as API references and FAQs. Particularly, they focus on advantages and disadvantages, highlighted features, and product comparison information in this stage. The questionnaire survey shows that comments on technical communities and GitHub projects related to the product are the two main factors that influence the technology selection process.

**The early stage**

Both junior and senior developers aim to get a general understanding of the concept, functions, and frameworks of the technical product in the early stage. The type of information, ways of accessing information, and user behaviors in this stage overlap largely with the exploration stage. The difference is that in this stage, the developers started to install and configure the environment, copy codes, and run a small demo to get started. The interaction between developers and information involves not only reading or watching but also active behaviors such as copying codes.

The main difference between junior and senior developers in this stage is that junior developers, especially beginning learners, prefer using textbooks and video tutorials as the primary source of information. Compared to official documentation, textbooks and video tutorials may be easier for these developers:

> *When I started to learn a new programming language, I prefer to read related textbooks or watch online courses first, because they provide concrete examples which is easier to understand.* (P6, Female, 1 year of experience)

**The middle stage**

The main action in the middle stage of learning is practice. Having grasped the basic knowledge of a technical product, the developers usually try to read and understand the sample codes. Then they try to implement small functions to be more familiar with the product. In this stage, sample codes provided by either colleagues or tutorials are an important source for developers. They frequently refer to the samples and step-by-step guide and modify the sample codes to implement simple functions.

> *I usually write down codes by myself to try to implement some simple*

*functions in my project. This process helps me to be familiar with the framework and be more proficient in programming.* (P22, Male, 3 years of experience)

It is noted that in the practice phase, some senior developers focus on their business needs. They aim to test the feasibility of the product to see if it could meet specific business needs. Therefore, they would conduct experiments on the product by modifying the source code, which was more detailed than the sample codes along with step instructions in the tutorials. One senior emphasized the importance of testing feasibility:

*What I care about is the stability and efficiency of the technical product. Before applying it to our projects, we would test its feasibility.* (P24, Male, 10 years of experience)

**The later stage**

The final stage is the application of the technical product into real projects. Developers usually implement advanced functions related to business in this stage. API references and FAQs are two important document types that provide technical details in this stage. Compared to junior developers, senior developers emphasize the role of source code because they want to know about the underlying logic of the technology.

Bug fixing and problem-solving are the most frequent actions in this stage. Most interviewees said that viewing official documentation was not the fastest way to find their desired answers. One developer explained the reasons:

*I don't like referring to official documentation because there may not be such specific information in the documentation. I mean, the problems or bugs we meet are so detailed and diverse that the documentation can't cover all of them. Instead, your problems may have been encountered by others, so it is very likely to find solutions on search engines.* (P8, Male, five years of experience)

What should be noted is that the problem-solving process is not linear but iterative. Developers may jump between different platforms and pages to find the right answer.

**Developers' expectations for documentation**

By analyzing the interview scripts regarding developers' expectations for quality documentation, we concluded five dimensions regarding the design of documentation, which were content, organization, maintenance (up-to-dateness), interaction design, and internationalization. The codes of the five categories are shown in Table 3.

**Table 3. Dimensions of Documentation Experience Design**

| Category | Code |
|---|---|
| Content | Clear language |
| | Explain the advantages and disadvantages of the product |
| | Provide technical background or explanation of principles |
| | Provide clear step-by-step instructions |
| | Provide examples |
| | Provide terminology explanation |
| | Provide illustrations |
| | Provide video tutorials |
|  | Provide FAQs or troubleshooting |
| Organization | Provide a clear structure |
|  | Content arranged from the easy to the difficult |
| Interaction design | Provide code online debugging |
|  | Provide two-way jump links |
|  | Provides inner search engines |
|  | Comfortable typography and foldable pages |
|  | Commands and codes can be copied with one click |
|  | Switchable page background-color |
|  | Documentation integrated into the IDE |
| Maintenance | Documents are updated timely |
|  | Provides multiple historical versions |
|  | Provide communication platforms (e.g. community) |
| Internationalization | Available in multiple languages |

The majority of interviewees considered that document content itself is the most important criterion of good developer documentation. The developers, especially junior developers, expected that the documentation provides not only step instructions but also explanations for principles. One participant stated:

> *A good document should not only tell me what to do and how to do something but also why I should do it like this.* (P1, Male, 3 years of experience)

In addition, senior developers highlighted the importance of an introduction to product advantages and disadvantages. One participant said:

> *I hope the content is presented straightforwardly. I mean, it's better to tell me the pros and cons of the product at the very beginning of the documentation. I need to compare different products to choose the best one.*
> (P17, Male, 5 years of experience)

The up-to-dateness of documentation is also a major concern of developers. Some developers complained about the delay in documentation maintenance:

> *The major problem of poor documentation is that the documents are outdated or there is no document at all.* (P-5, Male, 2 years of experience)

> *Documents that do not update timely can be troublesome because they can mislead the developers. The mismatch between documents and software is always frustrating.* (P-20, Male, 6 years of experience)

## Quantitative results

### Characteristics of developers

We performed a descriptive analysis of the survey data regarding the basic information of the junior and senior developers. The basic information of the two groups is shown in Table 4.

**Table 4. Basic Information of the Two Groups**

| Group | Average age (year) | Average work experience (year) | Major work city | Average working hours per day | Average learning hours per day | Mostly used programming languages |
|---|---|---|---|---|---|---|
| Junior developers | 25 | 1 | Shenzhen | 8-10 hours | 1-2 hours | Python, Java, HTML/CSS |
| Senior developers | 32 | 8 | Beijing | 10-12 hours | 2-3 hours | C++, JAVA, RUST |

The interview results showed that developers learned new knowledge and technology in different ways. Figure 4 shows the proportion of different ways of learning among the respondents. The result indicated that reading official documentation is not the major way adopted by developers.

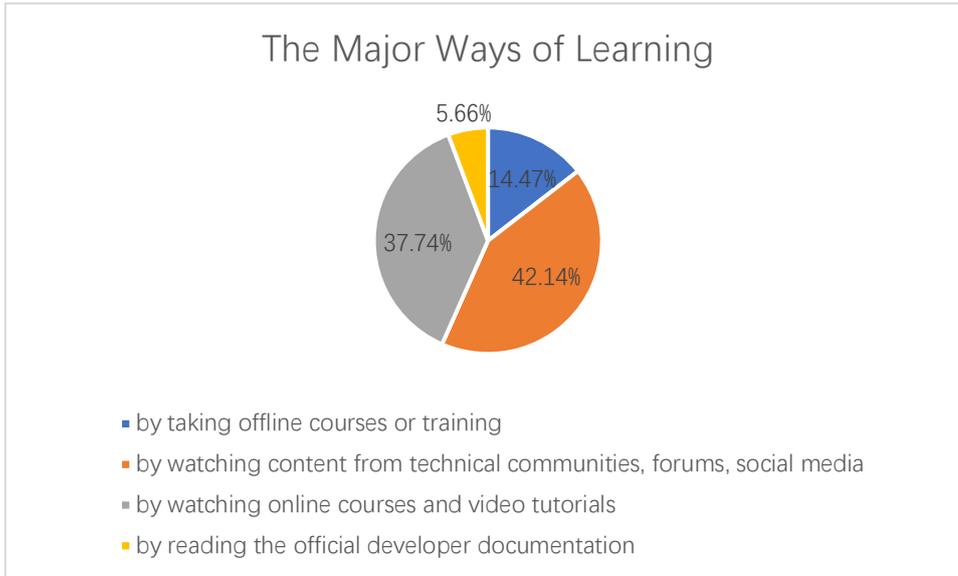

**Figure 4. The Major Ways of Learning**

Since the characteristics related to the learning habits of developers are more closely related to the use of technical documentation, in the survey, we used a five-point Likert scale (1=totally disagree, 5=totally agree) to examine whether these characteristics exist in the respondents. The results are shown in Table 5. It can be seen that the average scores of these characteristics are more than 3.00. Therefore, we think these characteristics are also present in the respondents.

**Table 5. The Average Score of Learning Habits**

| Description | Average score |
|---|---|
| I need to learn new knowledge and technology constantly and quickly. | 4.69 |
| I prefer to learn by myself. | 4.00 |
| The knowledge I learned is relatively fragmented. | 3.62 |
| I'm used to learning new techniques while working on projects. | 4.08 |

In terms of the preference for English or Chinese documentation, 53.84% of junior developers indicated that they prefer to read English documents directly, while only 30.76% of senior developers prefer to read English documents.

**Information journey of developers**

The interview results showed that the information journey of developers consists of the pre-learning stage, the early stage, the middle stage, and the later stage. We included the descriptions of the two draft maps in the survey. The result showed that the majority of participants (98.11%) agreed that the four stages can roughly describe their learning process, which indicated that the two drafts drawn from the interview data were correct. By comparing the qualitative codes and the data of the survey, we found that some codes that emerged from the interview were not represented in the survey. Therefore, we excluded those codes chosen by less than 30% of participants in the questionnaire survey and refined the two maps.

We found from the interviews that bug fixing and problem-solving were the most common actions in the later stage. Figure 5 shows various approaches that are employed to solve technical problems and their overall scoring.

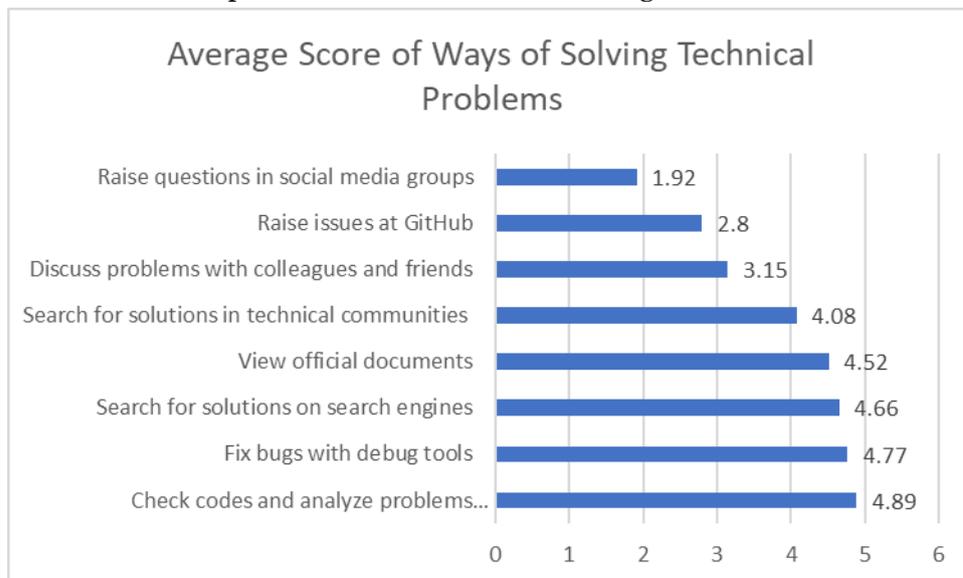

**Figure 5. Average Score of Ways of Solving Technical Problems**

It can be seen that viewing official documents are not the most common action taken by developers when they encounter technical problems. Analyzing problems independently (average rank score = 4.89), using debug tools (average rank score = 4.77), and search engines (average rank score = 4.66) are the top 3 common ways of solving problems. In addition, technical communities such as Stack Overflow are also popular knowledge sources for finding solutions.

**Developers' expectations for documentation**

We asked participants to rank the five dimensions in the questionnaire survey. The overall ranking of the five dimensions is shown in Figure 6.

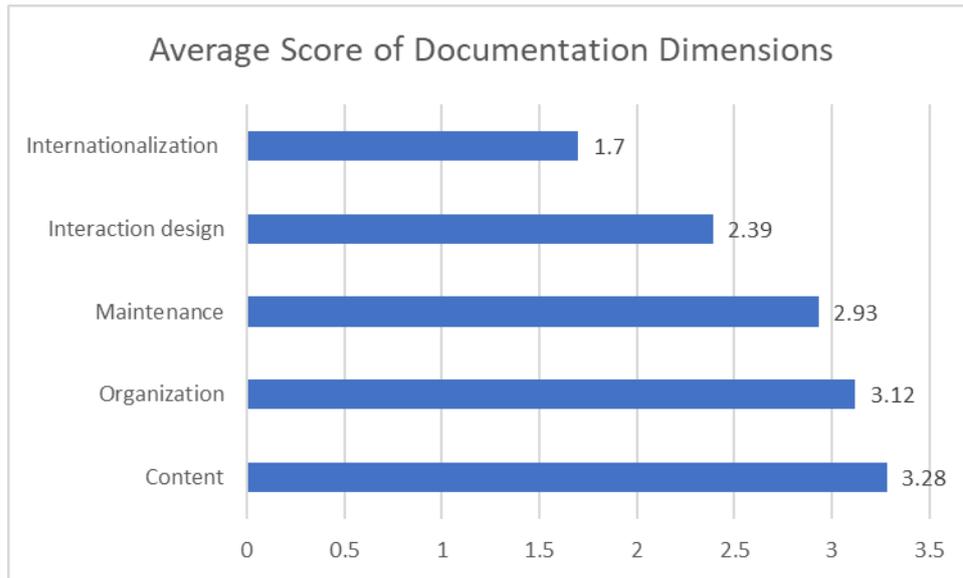

**Figure 6. Average Score of Documentation Dimensions**

The result of the survey is in line with the interview results. From Figure 6, it can be seen that developers value the content (average rank score = 3.28) of documentation most. The organization (average rank score = 3.12) and maintenance (average rank score = 2.93) of the documentation are the second and third most important aspects. Interaction design and internationalization of documentation, however, are less significant for the overall documentation experience.

## Discussion

### Personas

To make it easier for readers to use our research results, we processed quantitative and qualitative data into personas and user journey maps. The personas representing junior and senior developers were developed with a combination of the interview data and the survey data. The personas were demonstrated in Figure 7 and Figure 8.

The basic information in Table 4 was used in the personas. Males accounted for nearly 70% of junior developers and 87% of senior developers. To reflect the diversity, we chose a female from a medium-sized company (53%) to represent junior developers and a male from a large company (47%) to represent senior developers. The occupation with the highest proportion of junior developers was front-end engineer, and most of their professional backgrounds were software engineering. Developers holding a bachelor's degree account for 62% of junior developers, and 71.3% of senior developers held a master's degree or above. Therefore, the education levels of the two personas were bachelor and master respectively. Typical occupation among senior developers was algorithm engineer, and the majority of them majored in computer technology. The characteristics "open source spirit" and "ability to learn quickly" in Table 2 were included in the Bios of the personas. The names of the personas were randomly chosen from typical Chinese names, and the photos were generated using an online face generator.

We chose some typical features emerging from the interviews to differentiate junior and senior developers. For example, the interviews showed that junior developers liked to get started by watching online courses or video tutorials. We included a quote from one junior developer to indicate this feature. Compared to senior developers, junior developers preferred reading English documents, which was also reflected in the two personas. Senior developers were used to learning new technologies and solving problems by themselves. Their typical source of information was not videos, but official documents, content in technical communities, and academic papers. In addition, the interviews also showed that senior developers usually do a lot of research in the pre-learning stage, which was also reflected through a quote. The core needs of the developers were adapted from Table 3. The three core needs in the first persona were mainly put forward by junior developers in the interview, and the three ones in the second persona were from senior developers.

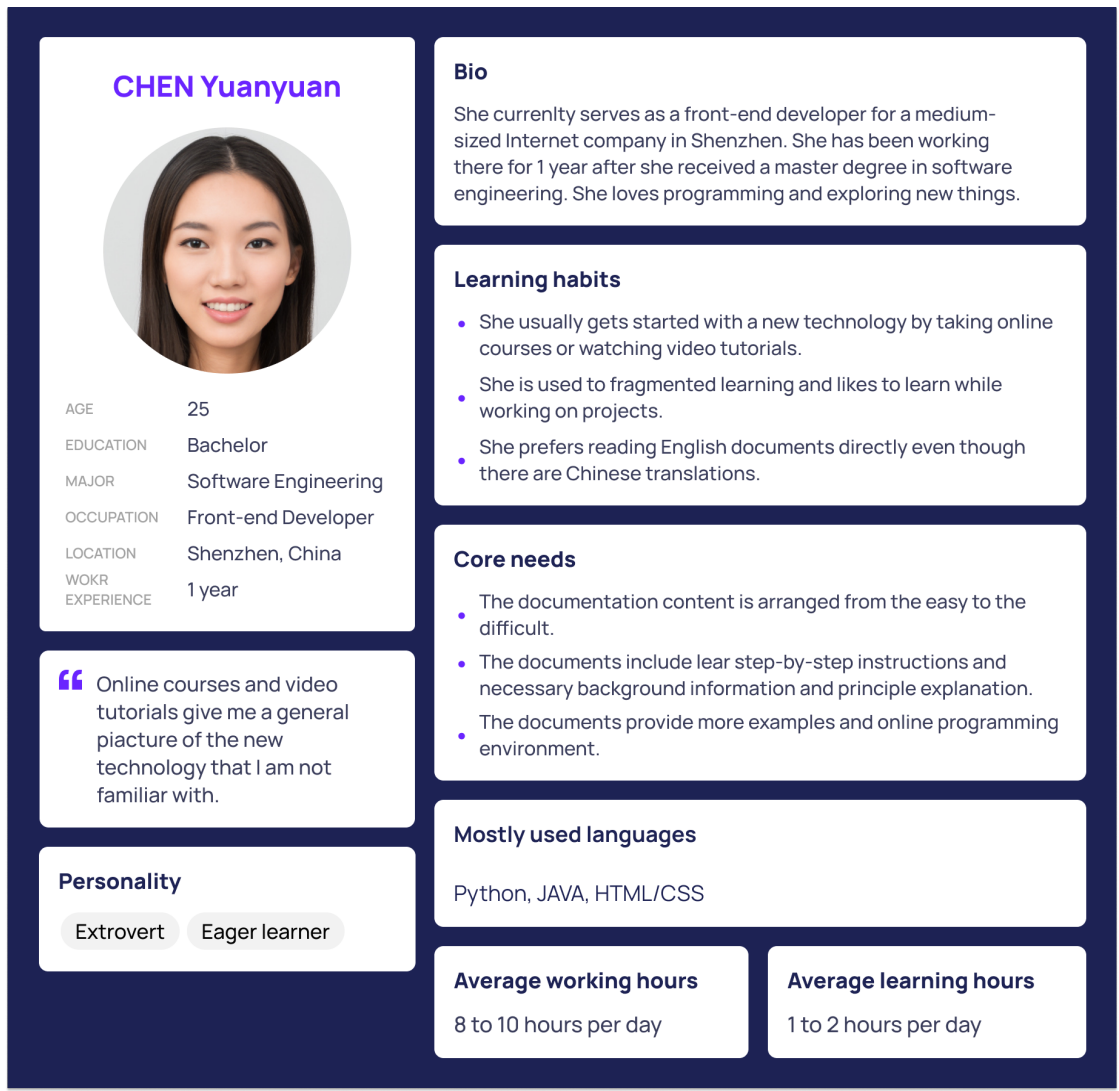

**Figure 7. The Persona of Junior Developers**

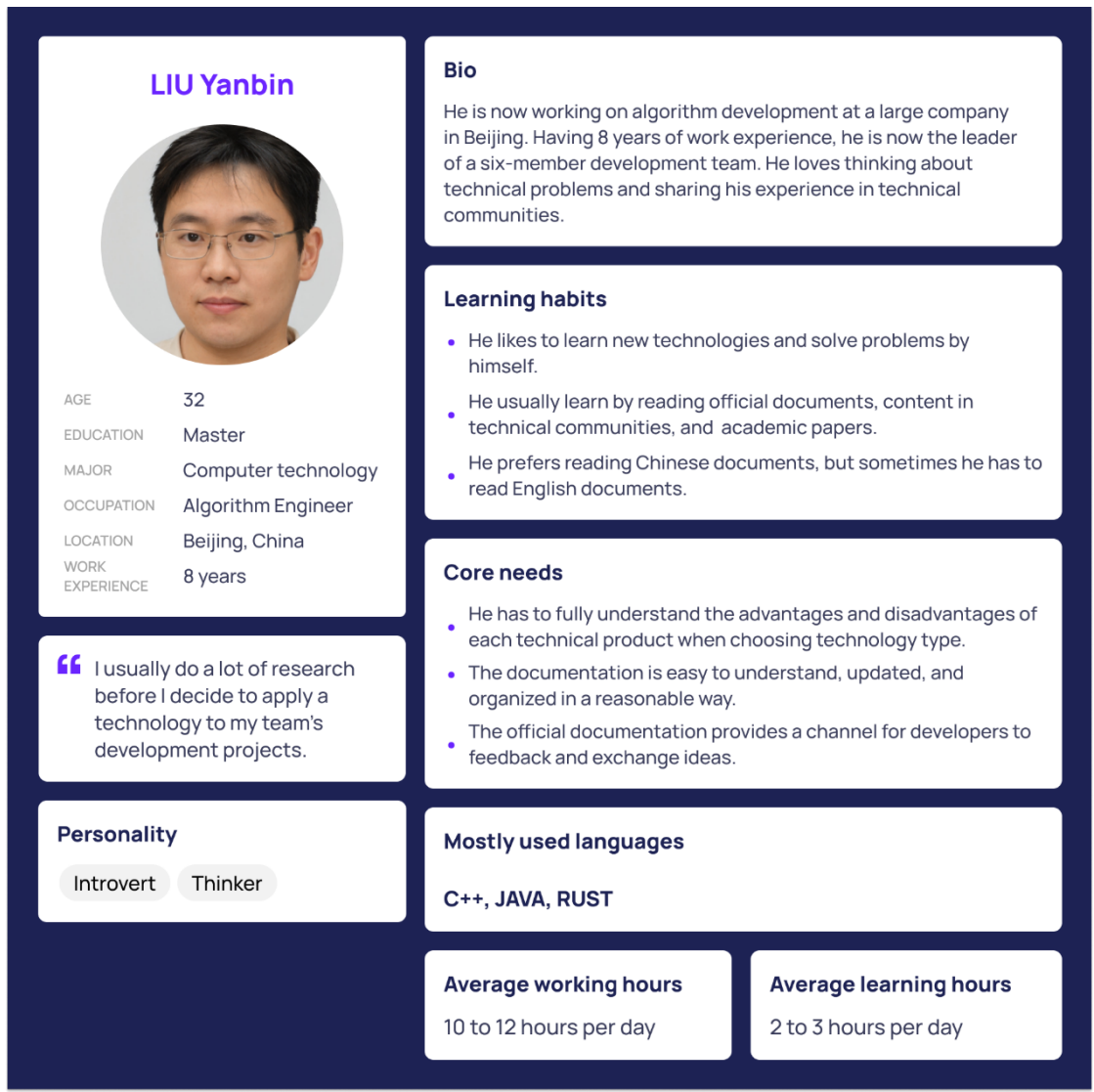

**Figure 8. The Persona of Senior Developers**

**Information journey maps**

As mentioned earlier, the drafts of the two information maps were refined by excluding those codes chosen by less than 30% of the respondents in the survey. To make the journey maps more concise, we combined those codes with similar meanings. For example, the original codes "implement advanced functions" and "improve existing functions" in the draft of the journey map of junior developers were combined into "implement advanced functions or improve existing functions". In addition, we changed the names of the four stages into the Exploration stage, the Understanding stage, the Practice stage, and the Application stage, because these four words can represent the goals and characteristics of the four stages respectively.

To help readers understand the journeys more concretely, we chose two typical scenarios from the experience of two interviewees. The scenario for the junior developer was to build a website for one of the products of her company. The scenario for the senior developer involved research, which required him to choose an appropriate open source machine learning framework for his project. The goals in these scenarios were also listed.

The two maps are shown in Figure 9 and Figure 10.

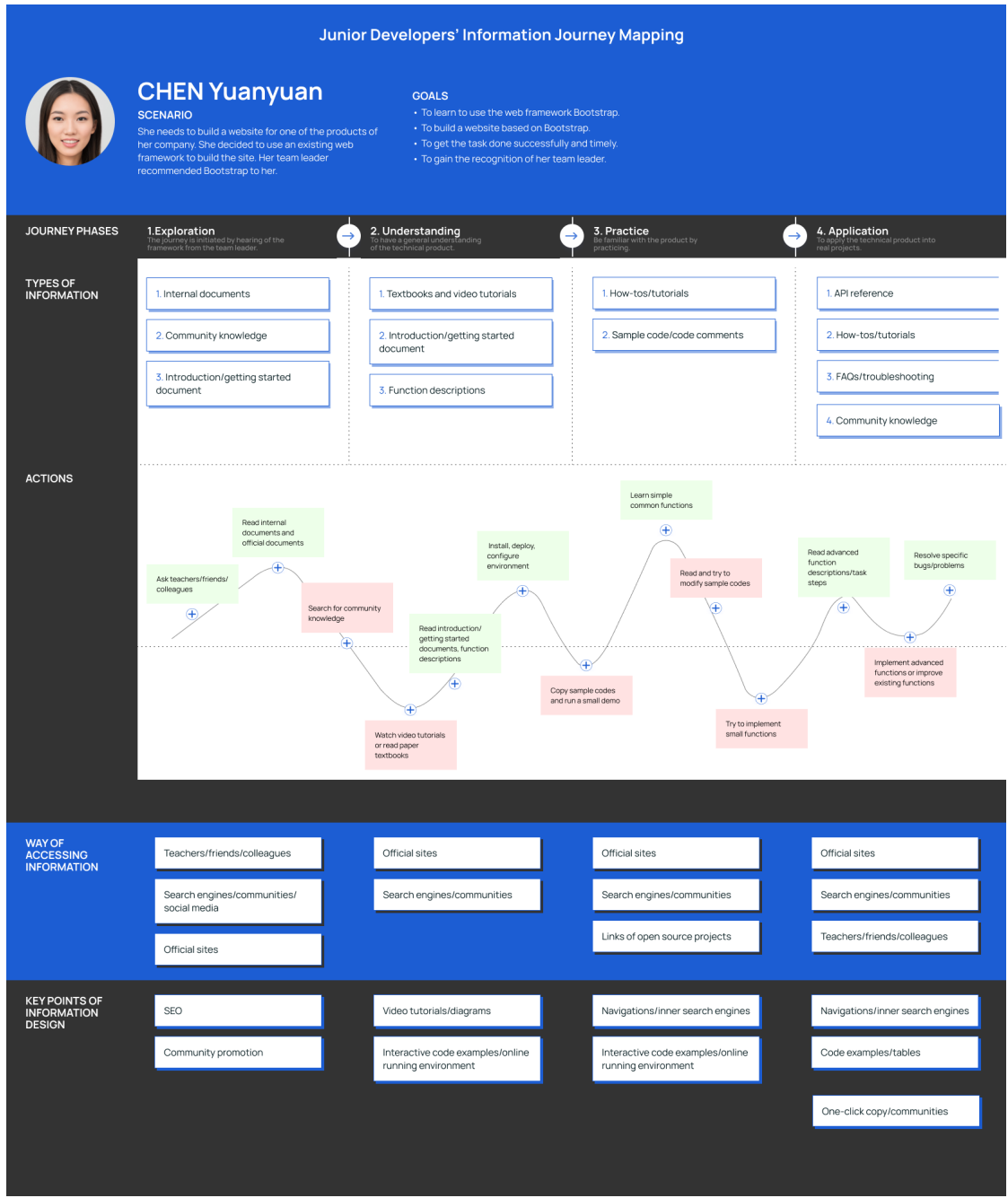

**Figure 9. Junior Developers' Information Journey Map**

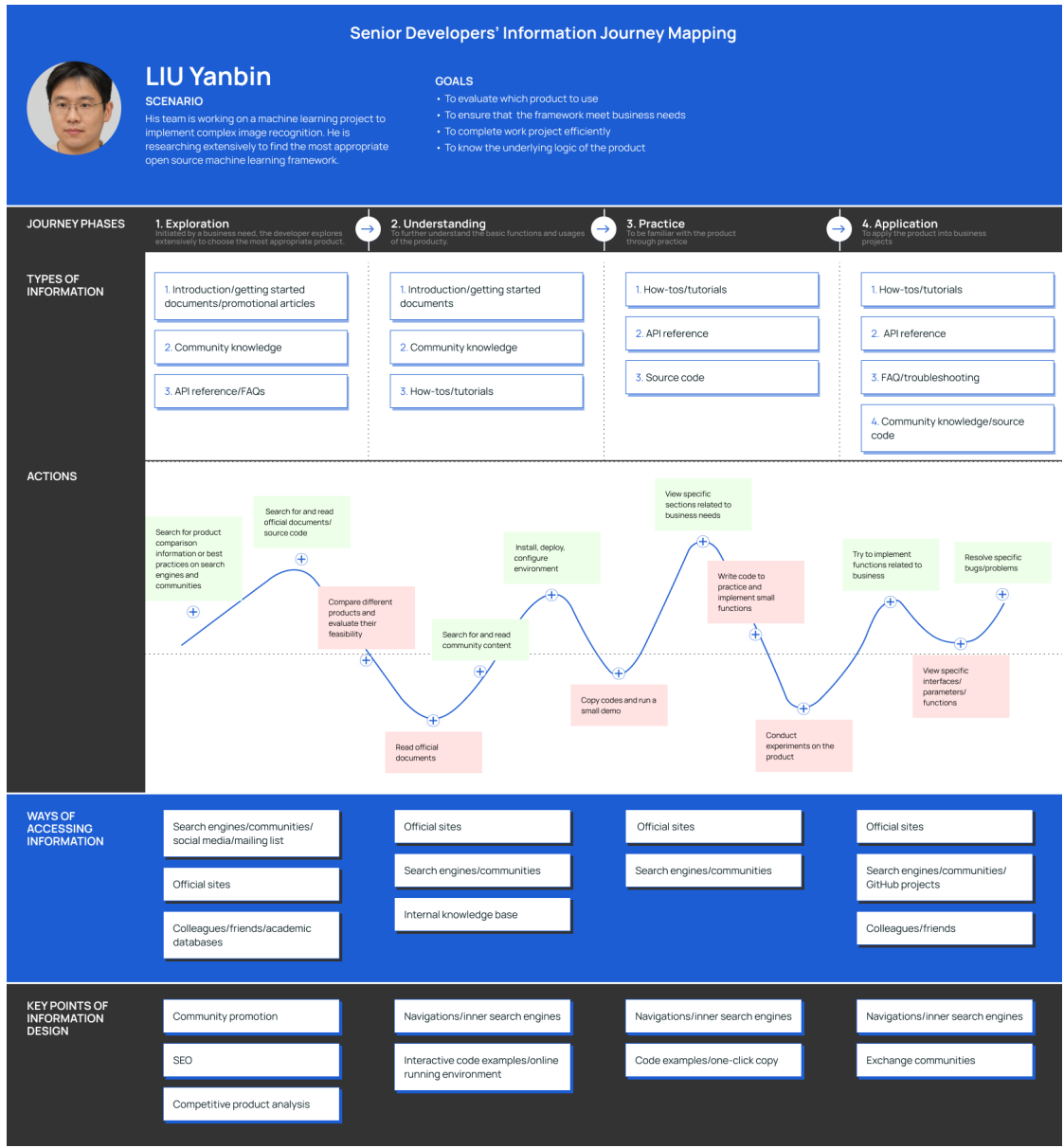

**Figure 10. Senior Developers' Information Journey Map**

**Implications for the design of developer documentation**

Our research is an initial attempt to explore the information journey of Chinese developers during the software development process. Our main findings are the two information journey maps of junior developers and senior developers who share some common patterns in their information behaviors. In the four stages of the information journey, the information types frequently used, the behaviors of developers, and their ways of accessing information are different, which may have implications for the design of developer documentation. We recommend the following suggestions to help improve developers' experience with documentation.

**Organize documentation content based on developers' information journey.** In the exploration and understanding stages of the journey, the developers focus on conceptual information such as introduction, function descriptions, and experience sharing in technical communities. In the practice stage, they rely heavily on code examples or demos to further familiarize themselves with the technical product. When they enter the application stage, code examples remain important, and document types containing technical details are the main information needs of developers. For instance, developers have to frequently use API references to check interfaces and parameters. Given that developers focus on different information types at different stages of the information journey, the organization of documentation content should be in line with their needs and preferences. If there is a documentation structure that can meet the expectations and learning habits of most developers, then this structure can be used as a template, which will likely improve the efficiency of documentation design and writing. In addition, documentation designers should consider the finding that most developers like learning by doing. Examples are essential document elements that allow developers to practice. Code examples of different difficulty levels can be presented in different parts of the document. For example, in the introductory section, provide a simple demo that can be run immediately. While in the middle part, provide more specific and in-depth examples.

**Documentation should be able to adapt to the different learning needs of developers at different levels.** As the personas and the information journey maps indicate, junior and senior developers have different habits and needs in their learning processes. One typical feature is that junior developers with less development experience prefer to get started by watching video tutorials. Providing options for watching short video tutorials in the quick start section of the documentation may be helpful for junior developers. Another aspect relevant to this concern is that the document content should be clear and complete enough to remove the barriers to understanding. Some junior developers complained in the interviews that some documents are not user-friendly because they lack necessary background information and principle explanation that may not be familiar to them. Another finding from the interview indicated that Chinese developers have different preferences for the language of documentation. The tendency that some developers prefer to read English documents is related to the quality of translations. Poorly translated Chinese documents will not only affect developers' experience but also their confidence in the product. Providing multilingual versions of documents and ensuring the quality of translation can improve the reading speed and reading experience of developers.

**Prioritize the design of basic experience dimensions.** Our research reveals that developers have different priorities for the dimensions of the documentation experience. Content, organization, and maintenance are basic experience dimensions most valued by developers. The content ranks first among all documentation dimensions. Correctness, clearness, and completeness of content are the most fundamental criteria for quality documentation. In terms of documentation maintenance, some management measures can be taken to ensure that documents are updated promptly. For instance, synchronize the document development cycle and software development cycle. Beyond the basic experience dimensions, special efforts can be taken to improve the interactive experience of developer documentation. Participants in our study emphasized that some interactive design elements, such as an online code editor or IDE, are very helpful and convenient for them to get started and practice.

**Integrate marketing communication and technical communication.** For a technical product, the design of developer documentation is not only an important part of technical communication but also an indispensable link in marketing communication. In our study, senior developers with years of experience in the IT industry emphasized the importance of technical selection. In the exploration stage, they search for information about technical products from multiple sources, including search engines, technical communities, forums, social media, and mailing lists. Peer comments on these platforms have a significant impact on their final choices. Therefore, implementing SEO and promoting products in these channels can be useful for making official documents more findable. In addition, to solve specific bugs and problems, developers are used to searching for answers on search engines or technical communities. Documentation designers can consider adding a platform in the product's documentation center that allows developers to exchange and share knowledge about the product, such as a Q&A community, message boards, and comments. This not only allows developers to save time from searching for answers on search engines but also provides a platform to gather more developers and improve the adhesiveness of users.

## Conclusion and Future Work

This paper reported our exploration of Chinese developers' documentation experience from an information journey mapping perspective. First, we draw the typical personas of junior and senior developers in China. The two groups of participants have different learning habits during their development process. Second, we analyzed their information journey patterns. The information journey of both two groups of developers roughly follows these four stages: Exploration, Understanding, Practice, and Application. Developers have various preferences for information types and demonstrated different information behaviors in these stages. Compared to junior developers, senior developers have more active interactions with information in the exploration stage. Third, we summarized developers' definitions of good documentation and identified the priorities of five documentation dimensions. Based on the findings, we provided suggestions for improving developers' experience with technical documentation.

This study is exploratory and has some limitations. The small number of participants makes it difficult to dig into more specific characteristics of information journey patterns. The research findings also lack the support of empirical studies. In the future, we will investigate more information behaviors of developers in a larger population.